\documentclass{PoS}

\title{Solar Event Simulations using the HAWC Scaler System}

\ShortTitle{Solar Event Simulations with HAWC}

\author{\speaker{Olivia Enriquez-Rivera}$^a$, Alejandro Lara$^a$ and Rogelio Caballero-Lopez$^a$ for the HAWC Collaboration$^b$ \\
        \llap{$^a$}Instituto de Geofisica, Universidad Nacional Autonoma de Mexico, Mexico city, Mexico \\
        \llap{$^b$}For a complete author list, see
\href{http://www.hawc-observatory.org/collaboration/icrc2015.php}{www.hawc-observatory.org/collaboration/icrc2015.php}.\\
        Email: \email{oenriquez@geofisica.unam.mx}, \email{alara@geofisica.unam.mx}}
\abstract{The High Altitude Water Cherenkov (HAWC) Observatory is an air shower array located near the volcano Sierra Negra in Mexico. The observatory has a scaler system sensitive to low energy cosmic rays (the geomagnetic cutoff for the site is 8 GV) suitable for conducting studies of solar or heliospheric transients such as Ground Level Enhancements (GLEs) and Forbush decreases. In this work we present the simulation of the HAWC response to these phenomena. We computed HAWC effective areas for different array configurations (different selection of photomultiplier tubes per detector) relevant for Forbush decreases and GLEs.}

\FullConference{The 34th International Cosmic Ray Conference,\\
		30 July- 6 August, 2015\\
		The Hague, The Netherlands}

\usepackage{lineno}
\begin{document}
%\linenumbers
\section{INTRODUCTION}

There are two types of solar events related to Galactic Cosmic Rays (GCR) that have been studied using ground-based instruments: Forbush decreases and Ground Level Enhancements (GLE) of cosmic ray intensity >1 GeV. Forbush decreases are produced by large magnetic structures propagating in the interplanetary medium which block and deviate the charged GCR particles, resulting in a deficit in the GCR intensity at Earth.  GLEs, on the other hand, represent the high energy extension of Solar Energetic Particles (SEPs) which may be produced either in the lower solar corona during flare events, or by shocks produced by Interplanetary Coronal Mass Ejections. At present, the accelerating mechanism(s) that gives rise to SEPs and/or GLEs is still a subject of debate. Moreover, the exponential rollover (or exponential cutoff) of GLE spectra at very high energies remains an open question.

Neutron monitors distributed around the world at different cutoffs are useful for measuring the energy spectrum of GLEs. However, the information provided by these instruments is limited.  They are omnidirectional counters with no angular sensitivity.  They come in two different versions with different sizes and are located at many different altitudes, producing a sensor array with many different geomagnetic and atmospheric cutoffs or thresholds.  The problems are greatest when trying to measure the spectrum during the anisotropic phase of a GLE, when only a few stations register a signal.  The use of "non standard" GCR detectors, such as muon telescopes and air shower arrays like HAWC add a new dimension to the observations to address GLEs. Both air shower arrays and muon telescopes are directionally sensitive and can yield information about the GLE anisotropy.  When used in proximity to a neutron monitor, such an instrument, even in an omnidirectional mode, provides an independent measure of the intensity at a different rigidity threshold.  Two pairs of neutron monitors now provide a similar function, those at Mt. Washington and Durham, New Hampshire and the South Pole and Polar Bare monitors in Antarctica.  The combination of the Climax NM and Milagro TeV gamma-ray telescope provided such coverage for the GLEs of 1997 November 6 \cite{2003ApJ...588..557F}, 2001 April 15 and 2005 January 20 \cite{2009ApJ...700L.127A}.

\section{HAWC}
The High Altitude Water Cherenkov (HAWC) Observatory is an air shower detector located at 4,100 m a.s.l, N $18^{\circ} 59' 48''$, W  $97^{\circ} 18' 34''$.  Build in the skirts of Sierra Niegra, Puebla in Mexico it consists of 300 steel Cherenkov detectors 7.3 m diameter and 4.5 m deep. Each tank is filled with filtered water and the total detector comprises an extension of $22,000$ $m^2$. The construction of the observatory started in 2011 and it was inaugurated in March 2015. During its construction phases HAWC has been continuously acquiring data with a growing detector. The observatory has a high duty cycle ($>95 \%$) and an instantaneous field of view of $\sim$ 2 sr.

Every day HAWC detects millions of cosmic-ray and gamma-ray showers produced when a primary charged particle or gamma ray enters the Earth's atmosphere. Since secondary particles produced in air showers travel faster than the velocity of light in water, they produce Cherenkov radiation in the detectors and four photomultiplier tubes (PMT) installed in the bottom of each tank detect Cherenkov photons. Each tank has 1 high quantum efficiency PMT located in its center and three lateral PMTs.

HAWC data is collected by two data acquisition systems (DAQs). The main DAQ measures arrival times and time over thresholds of PMT pulses which allows for the reconstruction of the air shower arrival direction and energy of the primary particle. The secondary DAQ consists of a scaler system that registers each time the PMT is hit by $>1/4$ photoelectron charge. This simpler system allows to measure particles below the energies of reconstructable showers.

HAWC demonstrated its capability for detecting solar transient events. During the HAWC prototype phase (VAMOS), a Forbush decrease was registered.  VAMOS consisted of seven tanks \cite{Castillo:2013}. At present, HAWC has observed around 5 Forbush decreases. In this paper we perform simulations of Forbush decreases and GLEs using different subarrays of the detector (PMT combinations) to understand the response of the HAWC scaler system to solar related phenomena

\section{HAWC SCALER SIMULATION}

  We begin by simulating the development of the secondary particle cascade produced by primaries using CORSIKA \cite{1998cmcc.book.....H}. 20 million primary protons were simulated with energies between 5 GeV and 2000 TeV at zenith angles between $0^{\circ}$ and $75^{\circ}$. Next, simulation of the particle propagation through the tanks is performed using GEANT-4 \cite{Agostinelli:2002hh}.  Finally, the response of the electronics is simulated.  We considered in our simulations three hundred tanks with different numbers of PMTs corresponding to different subarrays of the detector. The effective area of the configurations are obtained by counting the hits in all PMTs. Figure 1 shows the effective area of the scaler system for the PMT sets indicated in the figure. As expected, the effective area scales with the number of PMTs used in the calculation.

\begin{figure}[h]
\centerline{\includegraphics[trim=0cm 6.5cm 0cm 6.5cm, clip=true, height=7cm]{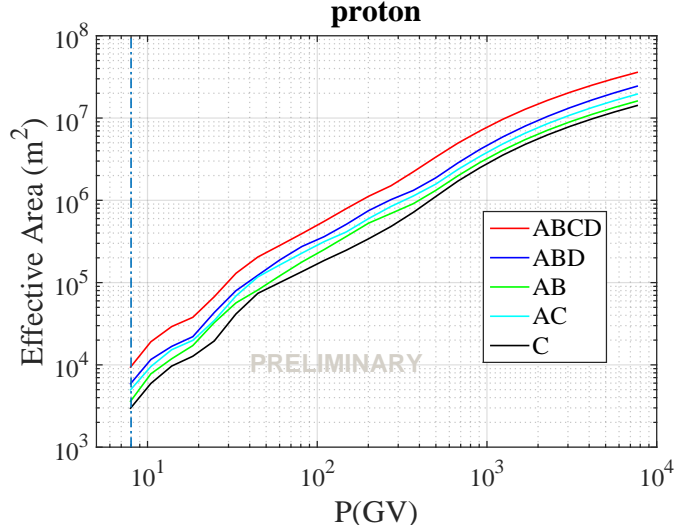}}
\caption{Effective areas for different subarrays of HAWC for protons thrown vertically. The blue vertical line indicates the geomagnetic energy cutoff of the site (8 GV). A,B and D represent responses from the non-central PMTs. C is the central PMT and A, B, D the surrounding PMTs in the tank. The central PMT has a higher quantum efficiency, which explains why the AB effective area is lower than AC. }
\end{figure}
  
To simulate the count rate $S$ of the detector, the convolution of $A_{eff}$ with the desired spectrum $J$ was computed. $S$ can be expressed as follows:

\begin{equation}
S(\theta)=\int_{P_{cut}}^{\infty}J(P,t)A_{Eff}(P,\theta) dP
\label{eq:eq1}
\end{equation}

where $P_{cut}$ is the geomagnetic cutoff rigidity, 8 GV.

\subsection{FORBUSH SIMULATION}
On 2014 September 14 HAWC registered a double-step Forbush decrease (Figure 2). The colours represent mean relative count rates of the HAWC scaler system, using different PMT combinations.

 The mean rate of each PMT configuration was obtained as follows: The mean counts per second are computed for each PMT in the array, then, we compute the mean rate in each set of PMT configurations (ABCD,ABD,C etc). Finally, taking as reference value (i.e. 100\%) the mean rate value observed during two days after the Forbush decrease (September 18 and 19), we compute the percentage of variation for each PMT configuration as seen in Figure 2. We took two days after the Forbush decrease because there was insufficient data prior to the Forbush decrease.

Below, we describe our simulations of the HAWC scaler system before the Forbush decrease and the decrease at its minimum (September 14). These are referred as $S_{bkg}$ and $S_{For}$ respectively.

\begin{figure}[h]
\centerline{\includegraphics[trim=1cm 0cm 0cm 0cm, clip=true, height=6cm]{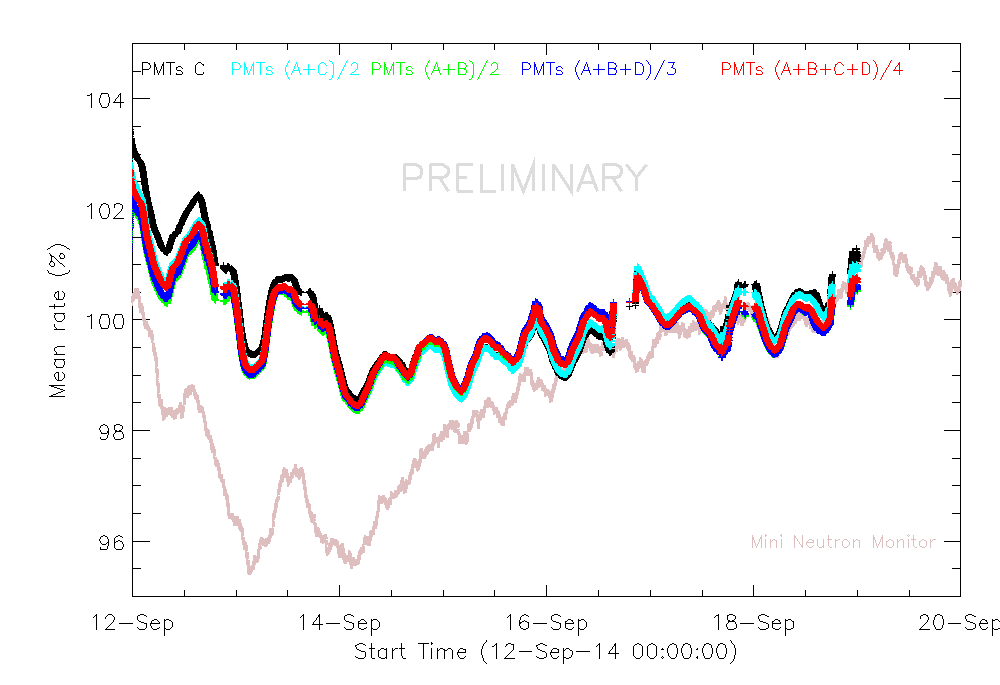}}
\caption{Forbush Decrease observed by HAWC and a mini neutron monitor located at HAWC's site on September 14, 2014. }
\end{figure}

For context, we show the force-field parameter $\phi$ from the force-field formalism developed for cosmic ray modulation in the heliosphere \cite{1968ApJ...154.1011G}. Based on the force-field approach, $\Delta \phi$ parameterizes the galactic spectrum $J_{gal}$ for different dates using the March 1987 spectrum as a reference ($\phi = 0$). We derived $\Delta \phi$ following the same procedure as described by \cite{2012JGRA..11712103C}, using neutron monitor data. For the March 1987 spectrum we used the same spectrum derived by \cite{2012JGRA..11712103C}.  This was obtained using measurements from PAMELA and IMP8 and fitting the data to the following double power law equation:

\begin{equation}
J(P)=J_{0}(P_{0}^a + P^a)^{(\gamma_1 - \gamma_2)/a}P^{\gamma_2}
\label{eq:eq2}
\end{equation}

The units of $J(P)$ and $J_{0}$ are $primaries/m^2\ s \ sr \ GV$. The parameters used to fit the proton spectrum from March 1987 and other spectra used in this work are presented in Table 1.

\begin{table}[h]
\centering

\begin{tabular}{c rrrrrrrrrrrrrrr}
\hline\hline
spectrum &solar cycle&$J_{0}$&$P_{0}$(GV) & a & $\gamma_1$ & $\gamma_2$ & Rig. Range &\\ [1ex]
%&&$primaries/m^2\ s \ sr \ GV$&&&&&\\ [1ex]
\hline % inserts single-line
%spectrum & $m^_{-2} s^_{-1} sr GV$\\ 
march 1987&minimum& 14000 & 0.96 & 1.5 & -2.7 & 2.8& $0.7<P<10^4$ \\
($\Delta\phi=0$ GV)&&&&&&&\\
sept 1989&maximum& 14000 & 3.1 & 1.7 & -2.7 & 1.2& $0.7<P<10^4$ \\
($\Delta\phi=0.55$ GV)&&&&&&&\\
sept 2014& maximum &14000 & 1.3 & 1.44 & -2.7 & 2.2& $0.7<P<10^4 $\\
($\Delta\phi=0.23$ GV)&&&&&&&\\
 sept 14, 2014 (FD)& maximum&14000 & 1.78 & 1.44 & -2.7 & 1.8& $0.7<P<10^4$ \\
($\Delta\phi=0.5$ GV)& & & & & & & \\[1ex]
\hline
\label{table:table1}
\end{tabular}
\caption{Parameters obtained to fit cosmic ray spectra used in this work with Eqt. 3.2. Units of $J_{0}$ are $primaries/m^2\ s \ sr \ GV$ }
\end{table}

To simulate the counts measured by HAWC before the Forbush decrease, $S_{bkg}$, the galactic cosmic ray spectrum from September 2014 was obtained using $\Delta \phi =0.23$ GV (see Table 1) and then convolved with the effective area (see equation 3.2). As pointed out by \cite{Usoskin:2015hoa}, the force-field approximation can be applied to obtain cosmic-ray spectra during Forbush decreases. The spectrum from the minimum of the Forbush decrease (September 14, 2014) was calculated using $\Delta\phi = 0.5$ GV and inserted in equation 3.2 to obtain $S_{For}$. For comparison, we computed the fractional decrease in counts $(S_{bkg}-S_{For})/S_{bkg}$ for our simulations and also for real HAWC data. These results are shown in Table 2.

\subsection{GLE SIMULATION}

We simulated the HAWC scaler response to GLE No. 42, 1989 September 29. This GLE was one of the largest in the past several decades, being detected by fifty-two neutron monitors and six muon telescopes, indicating a hard spectrum. To simulate the background counts observed by HAWC scalers we used the GCR spectrum $J_{gal}$ from September 1989 obtained with $\Delta \phi=0.55 GV$ (see Table 1). On the other hand, the GLE spectrum can be expressed as $J_{GLE}=J_{gal}+ J_s$ where $J_s$ represents the solar particle spectrum. To simulate the delayed component of GLE 42 we used a power law spectrum $J_s=J_0 E^{-\gamma}$ where $J_0=2.5 \cdot10^5$ and $\gamma=4$. This spectrum is close to other spectra reported in literature \cite{2011ASTRA...7..459V} \cite{2014ApJ...790..154M}. We insert $J_s$ in equation 3.1 to obtain the excess in expected counts observed by the HAWC scalers. In Table 2 we summarize the fractional increase of the GLE during its maximum peak obtained from the simulation. As indicated in Table 2, we obtained a maximum value of 44.4\% for PMT set ABCD. Due to the hard spectrum, the event was observed in several low-latitude neutron monitors, such as Mexico City, where the maximum peak registered an increase of 44.2\%.

\begin{table}[h]
\centering
\begin{tabular}{c cccccccccccccc}
\hline\hline
PMT CONFIG.& FD& FD &GLE \\ 
&OBS & SIM &SIM \\ [1ex]
&\%&\%&\%\\ [1ex]
\hline % inserts single-line
ABCD&2.85&2.10&44.4\\
ABD&2.83&2.00&41.1\\
AB&2.60&1.94&37.4\\
AC&2.80&2.10&43.4\\
C&2.91&2.08&42.8\\[1ex]
\hline
\end{tabular}
\caption{Fractional decreases (in percentage) obtained for the simulation of the Forbush decrease observed in September 14, and for GLE No. 42. The first column shows the actual fractional decreases observed by the HAWC scaler system during the Forbush event.}

\end{table}

\section{CONCLUSIONS}
HAWC is an air shower array sensitive to galactic and extragalactic gamma rays and cosmic rays. Due to its high altitude and location, it is well suited to perform studies of solar phenomena related to cosmic rays. HAWC has a large effective area and its design allows for different PMT configurations, or subarrays, representing different thresholds used in evaluating the spectrum of a transient. As a first step, we simulated two solar and heliospheric transients: GLEs and Forbush decreases. 

We simulated the fractional decrease during the FD from September 14, 2014.  The simulated rates are in agreement with HAWC data for several PMT combinations. For the GLE simulation, we could only compute the response to an archival event, and we found that it also reproduced the correct increase for the nearby Mexico City NM (rigidity cutoff=8.15 GV). Using a solar spectrum $J_s$ similar to spectrum reported in the literature for GLE 42, we found that HAWC would have easily registered GLE 42.  It is important to note that, as pointed out by \cite{2014ApJ...790..154M}, even for this event it is still difficult to calculate a precise spectrum for GLEs, i.e., we cannot distinguish between a power-law and exponential form at these energies. As a test, we tried different spectra $J$ with slight changes in $J_o$ and $\gamma$ (not shown here) and in some cases there were important variations in the expected counts. However, in all cases, HAWC was always able to detect GLE 42.  To see any exponential rollover, one must employ lower energy data from spacecraft as was done by \cite{2009ApJ...700L.127A}.

\section*{Acknowledgments}
\footnotesize{
We acknowledge the support from: the US National Science Foundation (NSF);
the US Department of Energy Office of High-Energy Physics;
the Laboratory Directed Research and Development (LDRD) program of
Los Alamos National Laboratory; Consejo Nacional de Ciencia y Tecnolog\'{\i}a (CONACyT),
Mexico (grants 260378, 55155, 105666, 122331, 132197, 167281, 167733);
Red de F\'{\i}sica de Altas Energ\'{\i}as, Mexico;
DGAPA-UNAM (grants IG100414-3, IN108713,  IN121309, IN115409, IN111315);
VIEP-BUAP (grant 161-EXC-2011);
the University of Wisconsin Alumni Research Foundation;
the Institute of Geophysics, Planetary Physics, and Signatures at Los Alamos National Laboratory;
the Luc Binette Foundation UNAM Postdoctoral Fellowship program.
}

\bibliography{icrc2015-0722}

\end{document}